\documentclass[11pt,draftcls,onecolumn]{IEEEtran}
\usepackage{times,amsmath,amssymb,dsfont,graphicx,float,enumerate}
\newtheorem{Theorem}{Theorem}
\newtheorem{Prop}{Proposition}
\newtheorem{Cor}{Corollary}
\newtheorem{Lem}{Lemma}
\newtheorem{Rem}{Remark}

\ifCLASSINFOpdf
\else
\fi
\IEEEoverridecommandlockouts

\begin{document}

\title{Learning in Hierarchical Social Networks}

\author{Zhenliang~Zhang,~\IEEEmembership{Student~Member,~IEEE,}
        Edwin~K.~P.~Chong,~\IEEEmembership{Fellow,~IEEE,}\\
        Ali~Pezeshki,~\IEEEmembership{Member,~IEEE,}
        William~Moran,~\IEEEmembership{Member,~IEEE,}
        and
        Stephen~D.~Howard,~\IEEEmembership{Member,~IEEE}
\thanks{This work was supported in part by AFOSR under Contract FA9550-09-1-0518, and by NSF under Grants ECCS-0700559, CCF-0916314, and CCF-1018472. Part of the material in this paper appears in preliminary conference papers \cite{SSP} and \cite{Allerton}.}
\thanks{Z.~Zhang, E.~K.~P.~Chong, and A.~Pezeshki are with the Department of Electrical and Computer Engineering, Colorado State University, Fort Collins, CO 80523-1373, USA (e-mail: zhenliang.zhang@colostate.edu; edwin.chong@colostate.edu; ali.pezeshki@colostate.edu).
}
\thanks{W.~Moran is with the Department of Electrical and Electronic Engineering, The University of Melbourne, Melbourne, VIC 3010, Australia (e-mail: wmoran@unimelb.edu.au).
}

\thanks{S.~D.~Howard is with the Defence Science and Technology Organisation, P.O. Box 1500, Edinburgh, SA 5111, Australia (e-mail: sdhoward@unimelb.edu.au).}
}

\maketitle

\begin{abstract}
We study a social network consisting of agents organized as a hierarchical $M$-ary rooted tree, common in enterprise and military organizational structures. The goal is to aggregate information to solve a binary hypothesis testing problem. Each agent at a leaf of the tree, and only such an agent, makes a direct measurement of the underlying true hypothesis. The leaf agent then generates a message and sends it to its supervising agent, at the next level of the tree. Each supervising agent aggregates the messages from the $M$ members of its group, produces a summary message, and sends it to its supervisor at the next level, and so on. Ultimately, the agent at the root of the tree makes an overall decision. We derive upper and lower bounds for the Type I and Type II error probabilities associated with this decision with respect to the number of leaf agents, which in turn characterize the converge rates of the Type I, Type II, and total error probabilities. We also provide a message-passing scheme involving non-binary message alphabets and characterize the exponent of the error probability with respect to the message alphabet size.
\end{abstract}


\section{Introduction}
We consider a binary hypothesis testing problem and an associated social network that attempts (jointly) to solve the problem. The network consists of a set of agents with interconnections among them. Each of the agents makes a measurement of the underlying true hypothesis, observes the past actions of his neighboring agents, and makes a decision to optimize an objective function (e.g., probability of error).
In this paper, we are interested in the following questions: Will the agents \emph{asymptotically learn} the underlying true hypothesis? More specifically, will the overall network decision converges in probability to the correct decision as the network size (number of agents) increases? If so, how fast is the convergence with respect to the network size? In general, the answers to these questions depend on the social network structure. 
There are two structures primarily studied in the previous literature.
\begin{itemize}
\item Feedforward structure: Each agent makes a decision sequentially based on its private measurement and the decisions of some or all previous agents. For example, we usually decide on which restaurant to dine in or which movie to go to based on our own taste and how popular they appear to be with previous patrons. Investors often behave similarly in asset markets.

\item Hierarchical tree structure: Each agent makes a decision based on its private measurement and the decisions of its descendent agents in the tree. This structure is common in enterprises, military hierarchies, political structures, online social networks, and even engineering systems (e.g., sensor networks).

\end{itemize}

The problem of social learning as described above is closely related to the decentralized detection problem. The latter concerns decision making in a sensor network, where each of the sensors is allowed to transmit a summarized message of its measurement (using a compression function) to an overall decision maker (usually called the fusion center). The goal typically is to characterize the optimal compression functions such that the error probability associated with the detection decision at the fusion center is minimized. However, this problem becomes intractable as the network structure gets complicated. Much of the recent work studies the decentralized detection problems in the asymptotic regime, focusing on the problems of the convergence and convergence rate of the error probability.

\subsection{Related Work}
The literature on social learning is vast spanning various disciplines including signal processing, game theory, information theory, economics, biology, physics, computer science, and statistics. 
Here we only review the relevant asymptotic learning results in the two aforementioned network structures.
\subsubsection{Feedforward Structure}
Suppose that a set of agents make decisions sequentially about the underlying truth $\theta$, which equals one of two hypotheses. The first agent makes a measurement of $\theta$ and generates a binary decision $d_1$, which is observed by all the other agents. The second agent makes its decision $d_2$ based on its own measurement and $d_1$. Recursively, the decision $d_N$ of the $N$th agent is based on its own measurement and the decisions observed from agents 1 to $N-1$. Banerjee \cite{Banerjee} and Bikchandani \emph{et al.} \cite{Bik} show that in the case where the agent signals only allow bounded private belief; i.e., the likelihood-ratio of each signal is bounded, if the first two agents make the same decision, then the rest of the agents would simply copy this decision ignoring their own measurements, even if their own measurements indicate the opposite hypothesis. This interesting phenomenon is also known as \emph{herding}. Moreover, we have $\lim_{N\to\infty} \mathbb P(d_N=\theta)<1,$ which means that the agent decisions do not converge in probability to the underlying true hypothesis as the number of agents goes to infinity; i.e., the agents cannot learn asymptotically.
Smith and Sorensen \cite{Smith} show that if the agent signals allow unbounded private beliefs; i.e., the likelihood-ratio of each signal can be greater than any constant, then these agents learn asymptotically. In other words, the agent decisions converge in probability to the underlying true hypothesis: $\lim_{N\to\infty} \mathbb P(d_N=\theta)=1.$ Krishnamurthy \cite{vicram0}, \cite{vicram} studies this problem from the perspective of quickest time change detection.
A similar scenario where agents make decisions sequentially but each agent only observes the decision from its immediate previous agent (also known as \emph{tandem network}) is considered in \cite{Tang}--\nocite{Tum,athans,Venu}\cite{tandem}. 
Veeravalli \cite{Venu} shows that the error probability converges sub-exponentially with respect to the number $N$ of agents in the case where the private measurements are independent and identically Gaussian distributed. Tay \emph{et al.} \cite{tandem} show that the error probability in general converges sub-exponentially and derive a lower bound for the error probability.  Djuric and Wang \cite{Djuric} investigate the evolution of social belief in these structures. Lobel \emph{et al.} \cite{acemoglu1} derive an upper bound for the error probability in the feedforward structure where each agent observes a decision randomly from all the previous agents.

\subsubsection{Hierarchical Tree Structure}
In many relevant situations, the social network structure is very complicated, wherein each individual makes its decision not by learning from all the past agent decisions, but from only a subset of agents that are directly connected to this individual. For complex network structures, Jadbabaie \emph{et al.} \cite{Ali} study the social learning problem from a non-Bayesian perspective. Acemoglu \emph{et al.} \cite{acemoglu} provide some sufficient conditions for agents to learn asymptotically from a Bayesian perspective. Cattivelli and Sayed \cite{sayed} study this problem using a diffusion approach. However, analyzing the convergence rate on learning for complex structures remains largely open.

Recent studies suggest that social networks often exhibit hierarchical structures \cite{Jameson}--\nocite{Wasserman,Clauset,Maiya,ya,Luis,Ec,Gupte,Motter}\cite{Noh}. These structures naturally arise from the concept of social hierarchy, which has been observed and extensively studied in fish, birds, and mammals \cite{Jameson}. Hierarchical structures can also be observed in networks of human societies \cite{Wasserman}; for example, in enterprise organizations, military hierarchies, political structures \cite{Maiya}, and even online social networks \cite{Gupte}.

In the special case where the tree height is 1, this structure is usually referred as the \emph{star configuration} \cite{Choi}--\nocite{Tenney,Chair,Cham,dec,Tsi,Tsi1,Warren,Vis,Poor,Sah,chen1,Liu,chen,Kas,Hao,Mou}\cite{Chong}. With the assumption of (conditional) independence of the agent measurements, the error probability in the star configuration converges exponentially with respect to the number $N$ of agents.
Tree networks with bounded height (greater than 1) are considered in \cite{Tang1}--\nocite{Nolte,tree1,tree2,tree3,Pete,Alh,Will}\cite{Lin}. In a tree network, measurements are summarized by leaf agents into smaller messages and sent to their parent agents, each of which fuses all the messages it receives with its own measurement (if any) and then forwards the new message to its parent agent at the next level. This process takes place throughout the tree, culminating at the root where an overall decision is made. In this way, information from each agent is aggregated at the root via a multihop path. Note that the information is `degraded' along the path. Therefore, the convergence rate for tree networks cannot be better than that of the star configuration. More specifically,
under the Bayesian criterion, the error probability converges exponentially fast to $0$ with an error exponent that is worse than the one associated with the star configuration~\cite{tree3}.

The error probability convergence rate in trees with unbounded height was considered in \cite{Zhang} and~\cite{yash}. We study in \cite{Zhang} the error probability convergence rate in balanced binary relay trees, where each nonleaf agent in this tree has two child agents and all the leaf agents are at the same distance from the root. Hence, this situation represents the worst-case scenario in the sense that the minimum distance from the root to the leaves is the largest. We show that if each agent in the tree aggregates the messages from its child agents using the unit-threshold likelihood-ratio test, then we can derive tight upper and lower bounds for the total error probability at the root, which characterize the convergence rate of the total error probability.
Kanoria and Montanari \cite{yash} provide an upper bound for the convergence rate of the error probability in $M$-ary relay trees (directed trees where each nonleaf node has indegree $M$ and outdegree $1$), with any combination of fusion rules for all nonleaf agents. Their result gives an upper bound on the rate at which an agent can learn from others in a social network.
To elaborate further, the authors of~\cite{yash} provide the following upper bound for the convergence rate of the error probability $P_N$ with any combination of fusion rules:
\begin{align}
\log_2 P_N^{-1}=O(N^{\log_M \frac{M+1}{2}}).
\label{eq}
\end{align}
They also provide the following asymptotic lower bound for the convergence rate in the case of majority dominance rule with random tie-breaking:
\[
\log_2 P_N^{-1}=\Omega(N^{\log_M \lfloor \frac{M+1}{2}\rfloor}).
\]
In the case where $M$ is odd, the majority dominance rule achieves
the upper bound in~\eqref{eq}, which shows that the bound is the optimal convergence rate. However, in the case
where $M$ is even, there exists a gap between these two bounds
because of the floor function in the second bound. 
In this case, \cite{yash} leaves two questions open:
\begin{itemize}
\item[Q1.] Does the majority dominance rule achieve the upper bound in \eqref{eq}?
\item[Q2.] Do there exist other strategies that achieve the upper bound in \eqref{eq}?
\end{itemize}
In our paper, for the case where $M$ is even, we answer the first
question definitively by showing that the majority dominance rule
does \emph{not} achieve the upper bound in \eqref{eq}. For the
second question, we provide a strategy that is closer to achieving
the upper bound in \eqref{eq} than the majority dominance rule.

Our paper also differs from (and complements)~\cite{yash} in a
number of other ways. For example, our analysis also includes non-asymptotic
results. Moreover, we also consider the Bayesian likelihood-ratio
test\footnote{By the Bayesian likelihood-ratio test, we mean a likelihood-ratio test in which the threshold is given by the ratio of the prior probabilities.} 
(the fusion rule for \emph{Bayesian learning}) as an alternative fusion rule, not considered in~\cite{yash}.
These differences should become clear as we clarify the
contributions of this paper in the next section.

\subsection{Contributions}

In this paper, we consider the learning problem in social networks configured as $M$-ary relay trees. Each agent at the leaf level, and only such an agent, takes a direct measurement of the underlying truth and generates a message, which is sent to its parent agent. Each intermediate agent in the tree receives messages from its child agents and aggregates them into a new message, which is again sent to its parent agent at the next level. This process takes place at each nonleaf agent culminating at the root, where a final decision is made. In this way, the information from the leaf agents is aggregated into a summarized form at the decision maker at the root. This hierarchical structure is of interest because it represents the worst-case scenario in the sense that the leaf agents are maximally far away from the decision maker at the root. 

In the study of social networks, $M$-ary relay trees arise naturally. First, as pointed out before, many organizational structures are well described in this way. Also, it is well-known that many real-world social networks, including email networks \cite{Ebel} and the Internet \cite{Yook}, are \emph{scale-free networks}; i.e., the probability $P(\ell)$ that $\ell$ links are connected
to a agent is $P(\ell)\sim c\ell^{-\gamma}$, where $c$ is a normalization constant and the parameter $\gamma\in (2,3)$. In other words, the number of  links does not depend on the network size and is bounded with high probability. Moreover, Newman \emph{et al.} \cite{Newman} show that the average degree in a social network is bounded or grows very slowly as the network size increases. Therefore, to study the learning problem in social networks, it is reasonable to assume that each nonleaf agent in the tree has a finite number of child agents, in which case the tree height grows unboundedly as the number of agents goes to infinity.

In this paper, we study two ways of aggregating information: the majority dominance rule (a typical non-Bayesian rule) and the Bayesian likelihood-ratio 
test. Our contributions are as
follows:
\begin{itemize}
\item[1)] In both cases, we have derived non-asymptotic bounds for the error probabilities with respect to the number of leaf nodes $N$. These bounds in turn characterize the asymptotic decay rates of the error probabilities.

\item[2)] Suppose that the majority dominance rule with random
tie-breaking is applied throughout the tree. In the case where $M$ is even, we derive the exact convergence rate of the error probability:
\[
\log_2 P_N^{-1}=\Theta(N^{\log_M \lfloor \frac{M+1}{2}\rfloor}).
\]
Therefore, we show that the majority dominance rule with random
tie-breaking does \emph{not} achieve the upper bound in \eqref{eq}.  (In the case where $M$
is odd, our asymptotic decay rate is consistent with the result in~\cite{yash}.) 
\item[3)] Suppose that the Bayesian likelihood-ratio tests is applied. We show that the convergence rate of the error probability is
\[
\log_2 P_N^{-1}=\Omega(N^{\log_M \lfloor \frac{M+1}{2}\rfloor}).
\]
Therefore, the convergence rate in this case is not worse than that in the majority dominance case. Hence in the case where $M$ is odd, the Bayesian likelihood-ratio test also achieves the upper bound in \eqref{eq}.

\item[4)] In the case where $M$ is even, we study an alternative majority dominance strategy, which achieves a strictly faster convergence rate than the majority dominance rule with random tie-breaking. The convergence rate of the total error probability using this strategy is 
\[
\log_2 P_N^{-1}=\Theta(N^{\log_M \sqrt{M(M+2)}/2}).
\]
The upper bound in \eqref{eq} involves an \emph{arithmetic} mean
of $M+2$ and $M$. In contrast, the above rate involves the
\emph{geometric} mean of $M+2$ and $M$. Therefore, the gap between
this rate and the upper bound in \eqref{eq}
is small and almost negligible when $M$ is large. 
We also show that the Bayesian likelihood-ratio test achieves this convergence rate under certain conditions.

\item[5)] We propose a message-passing scheme involving non-binary
message alphabets. We derive explicit convergence rates of the
total error probabilities in the following cases: any combination of
fusion rules, majority dominance rule with random tie-breaking,
Bayesian likelihood-ratio test, and alternative majority dominance
rule. We also derive tight upper and lower bounds for the average message size as explicit functions of the spanning factor $M$.

\end{itemize}


\section{Problem Formulation}
\label{section2}
\begin{figure}[htbp]
\centering
\includegraphics[width=3in]{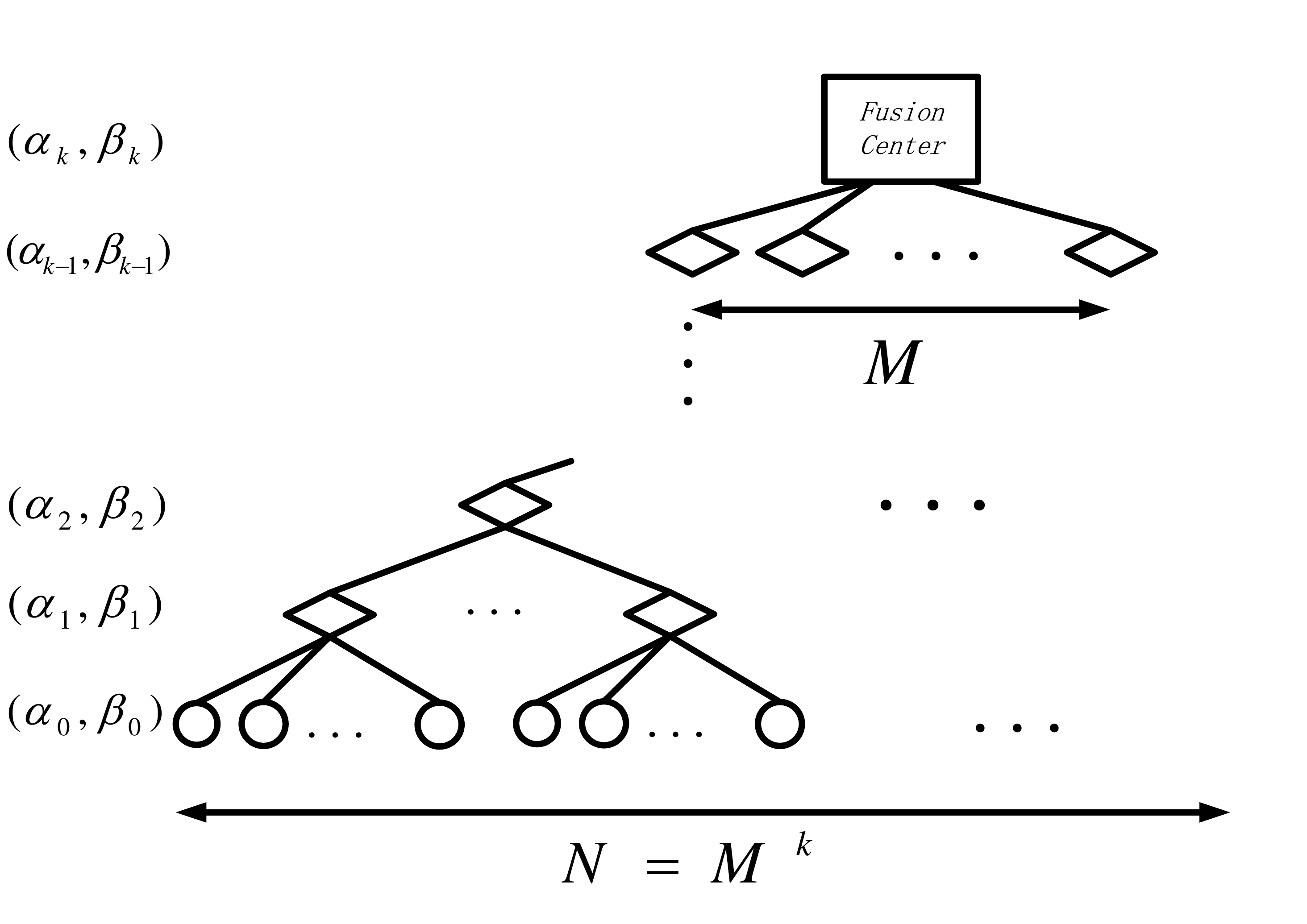}
\vspace{-0.3in}
\caption{An $M$-ary relay tree with height $k$. Circles represent leaf agents making direct measurements. Diamonds represent relay agents which fuse $M$ binary messages. The rectangle at the root makes an overall decision.}
\label{fig:tree}
\end{figure}

We consider the problem of binary hypothesis testing between $H_0$ and $H_1$, with $\mathbb{P}_0$ and $\mathbb{P}_1$ as the probability measures associated with the two hypotheses. The social network is organized as an $M$-ary relay tree shown in Fig. \ref{fig:tree}, in which leaf agents (circles) are agents making independent measurements of the underlying true hypothesis. Only these leaves have direct access to the measurements in the tree structure. These leaf agents then make binary decisions based on their measurements and forward their decisions (messages) to their parent agents at the next level. Each nonleaf agent, with the exception of the root, is a \emph{relay} agent (diamond), which aggregates $M$ binary messages received from its child agents into one new binary message and forwards it to its parent agent again. This process takes place at each agent, culminating at the root (rectangle) where the final decision is made between the two hypotheses based on the messages received. We denote the number of leaf agents by $N$, which also represents the number of measurements.
The height of the tree is $\log_M N$, which grows unboundedly as the number of leaf agents goes to infinity.

We assume that the decisions at all the leaf agents are independent given each hypothesis, and that they have identical Type I error probability (also known as false alarm probability, denoted by $\alpha_0$) and identical Type II error probability (also known as missed detection probability, denoted by $\beta_0$).
In this paper, we answer the following questions about the Type I and Type II error probabilities:
\begin{itemize}
\item How do they change as we move upward in the tree?
\item What are their explicit forms as functions of $N$?
\item Do they converge to $0$ at the root?
\item If yes, how fast will they converge with respect to $N$?
\end{itemize}
For each nonleaf agent, we consider two ways of aggregating $M$ binary messages:
\begin{itemize}
\item In the first case, each nonleaf agent simply aggregates $M$ binary messages into a new binary decision using the majority dominance rule (with random tie-breaking), which is a typical non-Bayesian fusion rule. This way of aggregating information is common in daily life (e.g., voting). For this fusion rule, we provide explicit recursions for the Type I and Type II error probabilities as we move towards the root. We derive bounds for the Type I, Type II, and total error probabilities at the root as explicit functions of $N$, which in turn characterize the convergence rates.

\item In the second case, each nonleaf agent knows the error probabilities associated with the binary messages received and it aggregates $M$ binary messages into a new binary decision using the Bayesian likelihood-ratio test, which is locally optimal in the sense that the total error probability after fusion is minimized. We derive an upper bound for the total error probability, which shows that the convergence speed of the total error probability using this fusion rule is at least as fast as that using the majority dominance rule.
\end{itemize}

\section{Error Probability Bounds and Asymptotic Convergence Rates: Majority Dominance}
\label{section3}

In this section, we consider the case where each nonleaf agent uses the majority dominance rule. We derive explicit upper and lower bounds for the Type I, Type II, and total error probabilities with respect to $N$. Then, we use these bounds to characterize the asymptotic convergence rates.

\subsection{Error Probability Bounds}

We divide our analysis into two cases: oddary tree ($M$ odd) and evenary tree ($M$ even). In each case, we first derive the recursions for the Type I and Type II error probabilities and show that all agents at level $k$ have the same error probability pair $(\alpha_k,\beta_k)$. Then, we study the step-wise reduction of each kind of error probability. From these we derive upper and lower bounds for the Type I, Type II, and the total error probability at the root.

\subsubsection{Oddary Tree}

We first study the case where the degree of branching $M$ is an odd integer. Consider an agent at level $k$, which aggregates $M$ binary messages $\textbf{u}_i^{k-1}=\{u_1^{k-1},u_2^{k-1},\ldots, u_M^{k-1}\}$ from its child agents at level $k-1$, where $u_t^{k-1}\in \{0,1\}$ for all $t$. Suppose that $u_o^k$ is the output binary message after fusion, which is again sent to the parent agent at the next level. The majority dominance rule, when $M$ is odd, is simply
\begin{equation*}
u_o^{k}:=\left\{\begin{array}{l l}
1, \quad  \text{ if }  \sum_{t=1}^{M} u_t^{k-1} \geq M/2,\\
0, \quad  \text{ if } \sum_{t=1}^{M} u_t^{k-1} \leq M/2. \end{array}\right.
\end{equation*}

Suppose that the binary messages $\{u_t^{k-1}\}_{t=1}^{M}$ have identical Type I error probability $\alpha$ and identical Type II error probability $\beta$. Then, the Type I and Type II error probability pair $(\alpha',\beta')$ associated with the output binary message $u_o^k$ is given by:
\begin{align*}
\alpha'=\mathbb{P}_0(u_o^k=1)&=\prod_{t=1} ^M \mathbb{P}_0(u_t^{k-1}=1)
+{M \choose 1}\mathbb{P}_0(u_s^{k-1}=0) \prod_{t=1} ^{M-1} \mathbb{P}_0(u_t^{k-1}=1)+\ldots\\
&+{M \choose (M-1)/2}\prod_{s=1}^{(M-1)/2} \mathbb{P}_0(u_s^{k-1}=0)\prod_{t=1} ^{(M+1)/2} \mathbb{P}_0(u_t^{k-1}=1)\\
&=f(\alpha), 
\end{align*}
where $f(\alpha) :=\alpha^M+{M \choose 1}\alpha ^{M-1}(1-\alpha)+\ldots+{M \choose (M-1)/2}\alpha^{(M+1)/2}(1-\alpha)^{(M-1)/2}$ and
\begin{align*}
\beta'=\mathbb{P}_1(u_o^k=0)&=\prod_{t=1} ^M \mathbb{P}_1(u_t^{k-1}=0)+{M \choose 1}\mathbb{P}_1(u_s^{k-1}=1) \prod_{t=1} ^{M-1} \mathbb{P}_1(u_t^{k-1}=0)+\ldots\\
&+{M \choose (M-1)/2}\prod_{s=1} ^{(M+1)/2} \mathbb{P}_1(u_s^{k-1}=1)\prod_{t=1}^{(M-1)/2} \mathbb{P}_1(u_t^{k-1}=0)\\
&=f(\beta).
\end{align*}

We assume that all the binary messages from leaf agents have the same error probability pair $(\alpha_0,\beta_0)$. Hence, all agent decisions at level $1$ will have the same error probability pair after fusion: $(\alpha_1,\beta_1)=(f(\alpha_0),f(\beta_0))$. By induction, we have
\begin{equation*}
(\alpha_{k+1}, \beta_{k+1})=(f(\alpha_k), f(\beta_k)), \quad \quad k=0,1,\ldots, \log_M N -1,
\label{equ:rel}
\end{equation*}
where $(\alpha_k,\beta_k)$ represents the error probability pair for agents at the $k$th level of the tree.
Note that the recursions for $\alpha_k$ and $\beta_k$ are identical. Hence, it suffices to consider only the Type I error probability $\alpha_k$ in deriving the error probability bounds. Before proceeding, we provide the following lemma.
\begin{Lem}
Let $h_k^M(x)=x^k+{M \choose 1} x^{k-1}(1-x)+\ldots+{M \choose k}(1-x)^k,$ where $k$ and $M$ are integers. Suppose that $0<k< M$. Then, $h_k^M$ is a monotone decreasing function of $x\in (0,1)$. 
\end{Lem}
\begin{IEEEproof}
We use induction in $M$ to prove the claim. First we note that $h_0^M(x)=1$ for all $M$.
Suppose that $M=2$. Then, we have $h_1^2(x)=2-x$.
Suppose that $M=3$. Then, we have
$h^3_1(x)=3-2x \text{ and } h^3_2(x)=x^2-3x+3.$
Clearly, in these cases $h_k^M$ are monotone decreasing functions of $x\in(0,1)$.

Now suppose that $h_k^{j}$ are monotone decreasing functions of $x\in(0,1)$ for all $j=2,\ldots,m-1$ and $k=1,\ldots, j-1$. We wish to show that $h_k^{m}$ are monotone decreasing functions of $x\in(0,1)$ for all $k=1,\ldots, m-1$.
We know that the binomial coefficients satisfy
\begin{align*}
{m \choose i}&={m-1 \choose i-1}+{m-1 \choose i}\\
&=  {m-1 \choose i-1}+{m-2 \choose i-1} +{m-2 \choose i}=\ldots\\
&=  {m-1 \choose i-1} +{m-2 \choose i-1}+\ldots+{k \choose i-1} +{k \choose i}.
\end{align*}
We apply the above expansion for all the coefficients in $h_k^m(x)$:
\begin{alignat*}{4}
h_k^m(x)  =x^k & +{m \choose 1} x^{k-1}(1-x)+\ldots+{m \choose k}(1-x)^k\\
                   = x^k & +{k\choose 1} x^{k-1}(1-x) +\ldots +{k \choose k} (1-x)^k \\
                                                & + {k\choose 0}x^{k-1}(1-x)+\ldots +{k \choose k-1} (1-x)^k+\ldots\\
                             & +{m-1\choose 0} x^{k-1}(1-x) +\ldots +{m-1 \choose k-1} (1-x)^k \\
                             =1&+ (1-x)h_{k-1}^k(x) +\ldots+ (1-x)h_{k-1}^{m-1}(x) \\
                             =1&+(1-x) \sum_{j=k}^{m-1} h_{k-1}^j(x).
\end{alignat*}
By the induction hypothesis, $h_{k-1}^{j}$ are monotone decreasing for all $j=k,\ldots, m-1$. Moreover, it is easy to see that $h_{k-1}^{j}$ are positive for all $j=k,\ldots, m-1$. Therefore, because the product of two positive monotone decreasing functions is also monotone decreasing, $h_k^m$ is a monotone decreasing function of $x\in(0,1)$. This completes the proof.
\end{IEEEproof}

Next we will analyze the step-wise shrinkage of the Type I error probability after each fusion step. This analysis will in turn provide upper and lower bounds for the Type I error probability at the root.

\begin{Prop}
Consider an $M$-ary relay tree, where $M$ is an odd integer. Suppose that we apply the majority dominance rule as the fusion rule. Then, for all $k$ we have
\[
1 \leq\frac{\alpha_{k+1}}{\alpha_k^{(M+1)/2}}\leq {M\choose (M-1)/2}.
\]
\label{prop1}
\end{Prop}

\begin{IEEEproof} Consider the ratio of $\alpha_{k+1}$ and $\alpha_k^{(M+1)/2}$:
\begin{align*}
\frac{\alpha_{k+1}}{\alpha_k^{(M+1)/2}}&=\alpha_k^{(M-1)/2}+{M \choose 1} \alpha_k^{(M-3)/2}(1-\alpha_k)+\ldots+{M\choose (M-1)/2} (1-\alpha_k)^{(M-1)/2}.
\end{align*}
First, we derive the lower bound of the ratio. We know that
\begin{align*}
1&=(\alpha_k+1-\alpha_k)^{(M-1)/2}=\alpha_k^{(M-1)/2}+{(M-1)/2 \choose 1}\alpha_k^{(M-3)/2}(1-\alpha_k)\\
&+\ldots+{(M-1)/2\choose (M-1)/2} (1-\alpha_k)^{(M-1)/2}.
\end{align*}
Moreover, it is easy to see that
${M \choose k}\geq {(M-1)/2 \choose k}
$ for all $k=1,2,\ldots, (M-1)/2$.
Consequently, we have
${\alpha_{k+1}}/{\alpha_k^{(M+1)/2}}\geq 1.$
Next, we derive the upper bound of the ratio. 
By Lemma 1, we know that the ratio $\alpha_{k+1}/\alpha_k^{(M+1)/2}$ is monotone increasing as $\alpha_k \to 0$. Hence, we have
\[
\frac{\alpha_{k+1}}{\alpha_k^{(M+1)/2}}\leq {M \choose (M-1)/2}.
\]
\end{IEEEproof}

The bounds in Proposition \ref{prop1} hold for all $\alpha_k\in (0,1)$. Furthermore, the upper bound is achieved at the limit as $\alpha_k\to 0$; i.e.,
$\lim_{\alpha_k\to 0} {\alpha_{k+1}}/{\alpha_k^{(M+1)/2}} = {M\choose (M-1)/2}.$
Using the above proposition, we now derive upper and lower bounds for $\log_2 \alpha_k^{-1}$.

\begin{Theorem}
Consider an $M$-ary relay tree, where $M$ is an odd integer. Let $\lambda_M=(M+1)/2$. Suppose that we apply the majority dominance rule as the fusion rule. Then, for all $k$ we have
\[
\lambda_M^k \left(\log_2 \alpha_0^{-1}-\log_2 {M\choose \lambda_M}\right) \leq \log_2 \alpha_k^{-1} \leq \lambda_M^k \log_2 \alpha_0^{-1}.
\]
\label{thm1}
\end{Theorem}

\begin{IEEEproof} From the inequalities in Proposition \ref{prop1}, we have
$\alpha_{k+1}=c_k\alpha_k^{(M+1)/2}=c_k\alpha_k^{\lambda_M},$
where $c_k\in \left[1,{M\choose (M-1)/2}\right]$. From these we obtain
\[
\alpha_k=c_{k-1}c_{k-2}^{\lambda_M} \ldots c_0^{{\lambda_M}^{k-1}} \alpha_0^{{\lambda_M}^k},
\]
where $c_i\in \left[1,{M\choose (M-1)/2}\right]$ for all $i$, and
\begin{align*}
\log_2 \alpha_k^{-1} = &-\log_2 c_{k-1}- \lambda_M \log_2 c_{k-2} -\ldots-\lambda_M^{k-1} \log_2 c_0 + \lambda_M^{k} \log_2 \alpha_0^{-1}.
\end{align*}
Since $\log_2 c_i \in \left[0,\log_2{M\choose (M-1)/2}\right]$, we have
$\log_2 \alpha_k^{-1}\leq \lambda_M^{k} \log_2 \alpha_0^{-1}.$
Moreover, we obtain
\begin{align*}
\log_2 \alpha_k^{-1} \geq &-\log_2{M\choose (M-1)/2} -\lambda_M \log_2{M\choose (M-1)/2} -\ldots\\
  &-\lambda_M^{k-1} \log_2{M\choose (M-1)/2} + \lambda_M^{k} \log_2 \alpha_0^{-1} \\
 =&-\frac{\lambda_M^k-1}{\lambda_M-1} \log_2{M\choose (M-1)/2} +\lambda_M^k \log_2 \alpha_0^{-1}\geq \lambda_M^k \left(\log_2 \alpha_0^{-1}-\log_2 {M\choose (M-1)/2}\right) \\
=& \lambda_M^k \left(\log_2 \alpha_0^{-1}-\log_2 {M\choose \lambda_M}\right).
\end{align*}

\end{IEEEproof}

The bounds for $\log_2 \beta_k^{-1}$ are similar and they are omitted for brevity. Note that our result holds for all finite integer $k$. In addition, our approach provides explicit bounds for both Type~I and Type II error probabilities respectively. From the above results, we immediately obtain bounds at the root simply by substituting $k=\log_M N$ into the bounds in Theorem \ref{thm1}.

\begin{Cor} Let $P_{F,N}$ be the Type I error probability at the root of an $M$-ary relay tree, where $M$ is an odd integer. Suppose that we apply the majority dominance rule as the fusion rule. Then, we have
\begin{align*}
N^{\log_M \lambda_M} \left(\log_2 \alpha_0^{-1}-\log_2 {M\choose \lambda_M}\right) \leq
\log_2 P_{F,N}^{-1}
\leq N^{\log_M \lambda_M} \log_2 \alpha_0^{-1}.
\end{align*}
\label{cor1}
\end{Cor}

\subsubsection{Evenary Tree}

We now study the case where $M$ is an even integer and derive upper and lower bounds for the Type I error probabilities. The majority dominance rule in this case is
\begin{equation*}
u_o^{k}:=\left\{\begin{array}{l l}
1, & \quad  \text{ if }  \sum_{t=1}^{M} u_t^{k-1} > M/2,\\

1 \text{ w.p. } P_b, & \quad  \text{ if }  \sum_{t=1}^{M} u_t^{k-1} = M/2,\\

0 \text{ w.p. } 1-P_b, & \quad  \text{ if }  \sum_{t=1}^{M} u_t^{k-1} = M/2,\\

0, & \quad  \text{ if } \sum_{t=1}^{M} u_t^{k-1} < M/2, \end{array}\right.
\end{equation*}
where $P_b\in (0,1)$ denotes the Bernoulli parameter for tie-breaking. We first assume that the tie-breaking is fifty-fifty; i.e., $P_b=1/2$. We will show later that this assumption can be relaxed. The recursions for the Type I and Type II error probabilities are as follows:
\begin{align*}
\alpha_k=\mathbb{P}_0(u_o^k=1)&=\prod_{t=1} ^M \mathbb{P}_0(u_t^{k-1}=1)
+{M \choose 1}\mathbb{P}_0(u_s^{k-1}=0) \prod_{t=1} ^{M-1} \mathbb{P}_0(u_t^{k-1}=1)+\ldots\\
&+\frac{1}{2}{M \choose M/2}\prod_{s=1}^{M/2} \mathbb{P}_0(u_s^{k-1}=0)\prod_{t=1}^{M/2} \mathbb{P}_0(u_t^{k-1}=1) \\
&=g(\alpha_{k-1}),
\end{align*}
where $g(\alpha_{k-1}):=\alpha_{k-1}^M+{M \choose 1}\alpha_{k-1}^{M-1}(1-\alpha_{k-1})+\ldots+\frac{1}{2}{M \choose M/2}\alpha_{k-1}^{M/2}(1-\alpha_{k-1})^{M/2}$ and
\begin{align*}
\beta_k=\mathbb{P}_1(u_o^k=0)&=\prod_{t=1} ^M \mathbb{P}_1(u_t^{k-1}=0)
+{M \choose 1}\mathbb{P}_1(u_s^{k-1}=1) \prod_{t=1} ^{M-1} \mathbb{P}_1(u_t^{k-1}=0)+\ldots\\
&+\frac{1}{2}{M \choose M/2}\prod_{s=1}^{M/2} \mathbb{P}_1(u_s^{k-1}=1)\prod_{t=1}^{M/2} \mathbb{P}_1(u_t^{k-1}=0) \\
&=g(\beta_{k-1}).
\end{align*}

Next we study the step-wise reduction of each type of error probability when each nonleaf agent uses the majority dominance rule. Again it suffices to consider $\alpha_k$ since the recursions are the same.

\begin{Prop}
Consider an $M$-ary relay tree, where $M$ is an even integer. Suppose that we apply the majority dominance rule as the fusion rule. Then, for all $k$ we have
\[
1 \leq\frac{\alpha_{k+1}}{\alpha_k^{M/2}}\leq \frac{1}{2}{M\choose M/2}.
\]
\label{prop2}
\end{Prop}

The proof is given in Appendix A. The upper bound is achieved at the limit as $\alpha_k\to 0$; i.e.,
$\lim_{\alpha_k\to 0} {\alpha_{k+1}}/{\alpha_k^{M/2}} = {M\choose M/2}/2.
$

In deriving the above results, we assumed that the tie-breaking rule uses $P_b=1/2$. Suppose now that the tie is broken with Bernoulli distribution with some arbitrary probability $P_b\in (0,1)$. Then, it is easy to show that
\[
P_b \leq \frac{\alpha_{k+1}}{\alpha_k^{M/2}}\leq 2^{M}.
\]
The bounds above are not as tight as those in Proposition \ref{prop2}. However, the asymptotic convergence rates remain the same as we shall see later.

Next we derive upper and lower bounds for the Type~I error probability at each level $k$.

\begin{Theorem}
Consider an $M$-ary relay tree, where $M$ is an even integer. Let $\lambda_M=M/2$. Suppose that we apply the majority dominance rule as the fusion rule. Then, for all $k$ we have
\[
\lambda_M^k \left(\log_2 \alpha_0^{-1}-\log_2{M\choose \lambda_M}  \right) \leq \log_2 \alpha_k^{-1} \leq \lambda_M^k \log_2 \alpha_0^{-1}.
\]
\label{thm2}
\end{Theorem}
The proof is given in Appendix B. Similar to the oddary tree case, we can provide upper and lower bounds for the Type I error probability at the root.

\begin{Cor}
Let $P_{F,N}$ be the Type I error probability at the root of an $M$-ary relay tree, where $M$ is an even integer. Suppose that we apply the majority dominance rule as the fusion rule. Then, we have
\begin{align*}
N^{\log_M \lambda_M}  \left(\log_2 \alpha_0^{-1}-\log_2{M\choose \lambda_M} \right) \leq
 \log_2 P_{F,N}^{-1} \leq  N^{\log_M \lambda_M} \log_2 \alpha_0^{-1}.
\end{align*}
\label{cor2}
\end{Cor}

\begin{Rem}
Notice that the above result is only useful when $M\geq 4$. For the case where $M=2$ (balanced binary relay trees), we have $\alpha_{k+1}=\alpha_k^2+\alpha_k(1-\alpha_k)=\alpha_k$
and $\beta_{k+1}=\beta_k^2+\beta_k(1-\beta_k)=\beta_k;$
that is, the Type I and Type II error probabilities remain the same after fusing with the majority dominance rule. 
\end{Rem}
\begin{Rem} We have provided a detail analysis in \cite{Zhang} of the convergence rate of the total error probability in balanced binary relay trees ($M=2$) using the unit-threshold likelihood-ratio test at every nonleaf agent. We show explicit upper and lower bounds for the total error probability at the root as function of the number $N$ of leaf agents, which in turn characterizes the convergence rate $\sqrt N$.
Moreover, we show that the unit-threshold likelihood-ratio test, which is locally optimal, is close-to globally optimal in terms of the reduction in the total error probability (see \cite{Zhang3} for details).
\end{Rem}

\begin{Rem}
Notice that the bounds in Corollaries \ref{cor1} and \ref{cor2} have the same form. Therefore, the odd and even cases can be unified if we simply let $\lambda_M=\lfloor(M+1)/2\rfloor$.
\end{Rem}

In the next section, we use the bounds above to derive upper and lower bounds for the total error probability at the root in the majority dominance rule case.

\subsubsection{Total Error Probability Bounds}
In this section, we provide upper and lower bounds for the total error probability $P_N$ at the root.
Let $\pi_0$ and $\pi_1$ be the prior probabilities for the two underlying hypotheses. It is easy to see that
$P_N=\pi_0 P_{F,N}+\pi_1 P_{M,N},$
where $P_{F,N}$ and $P_{M,N}$ correspond to the Type I and Type II error probabilities at the root. With the bounds for each type of error probability in the case where the majority dominance rule is used, we provide bounds for the total error probability as follows.

\begin{Theorem}
Consider an $M$-ary relay tree, let $\lambda_m=\lfloor(M+1)/2\rfloor$. Suppose that we apply the majority dominance rule as the fusion rule. Then, we have
\begin{align*}
N^{\log_M \lambda_M} \left(\log_2 \max\{\alpha_0,\beta_0\}^{-1}-\log_2{M \choose \lambda_M}\right)  \leq 
\log_2 P_N^{-1}
\leq
N^{\log_M \lambda_M} (\pi_0\log_2 \alpha_0^{-1}+\pi_1\log_2 \beta_0^{-1}).
\end{align*}
\label{thm3}
\end{Theorem}
\begin{IEEEproof}
From the definition of $P_N$; that is,
$P_N=\pi_0 P_{F,N}+\pi_1 P_{M,N},$
we have $P_N\leq \max\{P_{F,N}, P_{M,N}\}.$
In addition, we know that $\alpha_k$ and $\beta_k$ have the same recursion. Therefore, the maximum between the Type I and Type II error probabilities at the root corresponds to the maximum at the leaf agents.
Hence, we have
$N^{\log_M \lambda_M} \left(\log_2 \max\{\alpha_0,\beta_0\}^{-1}-\log_2{M \choose \lambda_M}\right)  \leq
\log_2 P_N^{-1}.$

By the fact that $\log_2 x^{-1}$ is a convex function, we have
$\log_2 P_N^{-1}\leq (\pi_0 \log_2 P_{F,N}^{-1}+\pi_1 \log_2 P_{M,N}^{-1}).$
Therefore, we have $\log_2 P_N^{-1}
\leq
N^{\log_M \lambda_M} (\pi_0\log_2 \alpha_0^{-1}+\pi_1\log_2 \beta_0^{-1}).$
\end{IEEEproof}

These non-asymptotic results are useful. For example, if we want to know how many measurements are required such that $P_N\leq \epsilon$, the answer is simply to find the smallest $N$ that satisfies the inequality in Theorem \ref{thm3}; i.e.,
\[
N^{\log_M \lambda_M} \left(\log_2 \max\{\alpha_0,\beta_0\}^{-1}-\log_2{M \choose \lambda_M}\right) \geq
\log_2 \epsilon^{-1}.
\]
Hence we have
\[
N \geq \left(\frac{\log_2 \epsilon^{-1}}{\log_2 \max\{\alpha_0,\beta_0\}^{-1}-\log_2{M \choose \lambda_M}}\right)^{\log_{\lambda_M} M}.
\]
The growth rate for the number of measurements is $\Theta({(\log_2 \epsilon^{-1})}^{\log_{\lambda_M} M})$.

\subsection{Asymptotic Convergence Rates}

In this section, we study the convergence rates of error probabilities in the asymptotic regime as $N \to \infty$. We use the following notation to characterize the scaling law of the asymptotic decay rate. Let $j$ and $h$ be positive functions defined on positive integers. We write $j(N)=O(h(N))$ if there exists a positive constant $c_1$ such that $j(N)\leq c_1 h(N)$ for sufficiently large $N$. We write $j(N)=\Omega(h(N))$ if there exists a positive constant $c_2$ such that $j(N)\geq c_2 h(N)$ for sufficiently large $N$.  We write $j(N)=\Theta(h(N))$ if $j(N)=O(h(N))$ and $j(N)=\Omega(h(N))$. 
%

From Corollaries \ref{cor1} and \ref{cor2}, we can easily derive the decay rates of the Type I and Type II error probabilities. For example, for the Type I error probability, we have the following.

\begin{Prop} Consider an $M$-ary relay tree, let $\lambda_M=\lfloor(M+1)/2\rfloor$. Suppose that we apply the majority dominance rule as the fusion rule. Then, we have $\log_2 P_{F,N}^{-1} =\Theta(N^{\log_{M} \lambda_M}).$
\label{prop3}
\end{Prop}
\begin{IEEEproof}
To analyze the asymptotic rate, we may assume that $\alpha_0$ is sufficiently small. More specifically, we assume that $\alpha_0<1/{M\choose \lambda_M}$. In this case, the bounds in Corollaries \ref{cor1} and \ref{cor2} show that
$\log_2 P_{F,N}^{-1} =\Theta(N^{\log_{M} \lambda_M}).$
\end{IEEEproof}

\begin{Rem}
Note that $\log_{M} \lambda_M$ is monotone increasing with respect to $M$. Moreover, as $M$ goes to infinity, the limit of $\log_{M} \lambda_M$ is 1.
That is to say, when $M$ is very large, the decay is close to exponential, which is the rate for star configuration and bounded-height trees. In terms of tree structures, when $M$ is very large, the tree becomes short, and therefore achieves similar performance to that of bounded-height trees.
\end{Rem}
\begin{Rem}
From the fact that the Type I and Type II error probabilities follow the same recursion, it is easy to see that the Type II error probability at the root also decays to 0 with exponent $N^{\log_{M} \lambda_M}$. 
\end{Rem}

Next, we compute the decay rate of the total error probability.

\begin{Cor} Consider an $M$-ary relay tree, let $\lambda_M=\lfloor(M+1)/2\rfloor$. Suppose that we apply the majority dominance rule as the fusion rule. Then, we have
$\log_2 P_{N}^{-1} =\Theta(N^{\log_{M} \lambda_M}).$
\label{cor3}
\end{Cor}

For the total error probability at the root, we have similar arguments with that for individual error probabilities. For large $M$, the decay of the total error probability is close to exponential.

\section{Error Probability Bounds and Asymptotic Convergence Rates: Bayesian Likelihood-ratio test}
\label{section4}
In this section, we consider the case where the Bayesian likelihood-ratio test is used as the fusion rule. We derive an upper bound for the total error probability, which in turn characterizes the convergence rate. We show that the convergence rate in this case is at least as fast or faster than that with the majority dominance rule.

\begin{Theorem} Let $\mathds{P}_N$ be the total error probability at the root in the case where the Bayesian likelihood-ratio test is used as the fusion rule in $M$-ary relay trees. We have
\begin{align*}
\log_2 \mathds{P}_N^{-1} \geq N^{\log_M \lambda_M}  \left(\log_2 L_0^{-1}-\log_2 \left(\frac{2{M \choose \lambda_M}\max(\pi_0,\pi_1)}{\min(\pi_0,\pi_1)^{\lambda_M}}\right) \right).
\end{align*}
\label{thm4}
\end{Theorem}
\begin{IEEEproof}
In the case where the majority dominance rule is used, from Propositions \ref{prop1} and \ref{prop2}, it is easy to show that
\[
\frac{1}{2}\leq\frac{\alpha_{k+1}+\beta_{k+1}}{\alpha_k^{\lambda_M}+\beta_{k}^{\lambda_M}}\leq 2{M \choose \lambda_M}.
\]
Since $x^{\lambda_M}$ is a convex function for all $M\geq2$, we have
\[
\frac{\alpha_k^{\lambda_M}+\beta_{k}^{\lambda_M}}{2} \geq \left(\frac{\alpha_k+\beta_k}{2}\right)^{\lambda_M},
\]
which implies the following:
\[
2^{-\lambda_M+1}\leq\frac{\alpha_k^{\lambda_M}+\beta_{k}^{\lambda_M}}{(\alpha_k+\beta_k)^{\lambda_M}}\leq 1.
\]
Hence, we obtain
\[
2^{-\lambda_M}\leq\frac{\alpha_{k+1}+\beta_{k+1}}{(\alpha_k+\beta_k)^{\lambda_M}}\leq 2{M \choose \lambda_M}.
\]
From these bounds and the fact that $\min(\pi_0,\pi_1) (\alpha_k+\beta_k)\leq \pi_0 \alpha_k+\pi_1\beta_k \leq \max(\pi_0,\pi_1) (\alpha_k+\beta_k)$, we have
\[
\frac{2^{-\lambda_M} \min(\pi_0,\pi_1)}{\max(\pi_0,\pi_1)^{\lambda_M}}\leq\frac{\pi_0 \alpha_{k+1}+\pi_1 \beta_{k+1}}{(\pi_0 \alpha_k+\pi_1\beta_k)^{\lambda_M}}\leq \frac{2{M \choose \lambda_M}\max(\pi_0,\pi_1)}{\min(\pi_0,\pi_1)^{\lambda_M}}.
\]
Note that $\pi_0\alpha_k+\pi_1 \beta_k$ is the total error probability for agents at level $k$ and we denote it by $L_k$.

The Bayesian likelihood-ratio test is the optimal rule in the sense that the total error probability is minimized after fusion. Let $L_k^{LRT}$ be the total error probability after fusing with the Bayesian likelihood-ratio test. We have
\[
\frac{L_{k+1}^{LRT}}{L_k^{\lambda_M}}\leq \frac{L_{k+1}}{L_k^{\lambda_M}}\leq \frac{2{M \choose \lambda_M}\max(\pi_0,\pi_1)}{\min(\pi_0,\pi_1)^{\lambda_M}}.
\]
Using a similar approach as that used in proving Theorem \ref{thm1}, we can derive the following lower bound for $\log_2 \mathds{P}_N^{-1}$:
\begin{align*}
\log_2 \mathds{P}_N^{-1} \geq N^{\log_M \lambda_M}  \left(\log_2 L_0^{-1}-\log_2 \left(\frac{2{M \choose \lambda_M}\max(\pi_0,\pi_1)}{\min(\pi_0,\pi_1)^{\lambda_M}}\right) \right).
\end{align*}

\end{IEEEproof}

From the above bound, we immediately obtain the following.

\begin{Cor} Consider an $M$-ary relay tree, and let $\lambda_M=\lfloor(M+1)/2\rfloor$. Suppose that we apply the Bayesian likelihood-ratio test as the fusion rule. Then, we have
$\log_2 \mathds{P}_N^{-1} = \Omega(N^{\log_{M} \lambda_M}).
$\end{Cor}

Note that in the case where the majority dominance rule is used, the convergence rate is exactly $\Theta(N^{\log_{M} \lambda_M})$.
Therefore, the convergence rate for the Bayesian likelihood-ratio test is at least as good as that for the majority dominance rule.

\section{Asymptotic Optimality of Fusion Rules} 
\label{section5}

In this section, we discuss the asymptotic optimality of the two fusion rules considered in our paper by comparing our asymptotic convergence rates with those in \cite{yash}, in which it is shown that with any combination of fusion rules, the convergence rate is upper bounded as
\begin{equation}
\log_2 P_N^{-1}=O(N^{\log_M \frac{(M+1)}{2}}).
\label{equ:upper}
\end{equation}

\subsection{Oddary case}
In the oddary tree case, if each nonleaf agent uses the majority dominance rule, then the upper bound in \eqref{equ:upper} is achieved; i.e.,
\begin{equation*}
\log_2 P_N^{-1}=\Theta(N^{\log_M \lfloor\frac{(M+1)}{2}\rfloor})=\Theta(N^{\log_M \frac{M+1}{2}}).
\end{equation*}
This result is also mentioned in \cite{yash}.
Tay \emph{et al.} \cite{tree3} find a similar result in bounded-height trees; that is, if the degree of branching for all the agents except those at level~1 is an odd constant, then the majority dominance rule achieves the optimal exponent.

Now we consider the case where each nonleaf agent uses the Bayesian likelihood-ratio test. Since the convergence rate for this fusion rule is at least as good as that for the majority dominance rule, it is evident that the Bayesian likelihood-ratio test, which is only locally optimal (the total error probability after each fusion is minimized), achieves the globally optimal convergence rate.  This result is also of interest in decentralized detection problems, in which the objective is usually to find the globally optimal strategy. In oddary trees, the myopically optimal Bayesian likelihood-ratio test, which is relevant to social learning problems because of the selfishness of agents, is essentially globally optimal in terms of achieving the optimal exponent.

\begin{Rem} Suppose that each nonleaf agent uses the Bayesian likelihood-ratio test and we assume that the two hypotheses are equally likely. In this case, the output message is give by the unit-threshold likelihood-ratio test:
\[
\frac{\prod_{t=1} ^M \mathbb{P}_1(u_t^{k-1})}{\prod_{t=1} ^M \mathbb{P}_0(u_t^{k-1})} \mathop{\gtrless}\limits_{H_{0}}^{H_{1}} 1.
\]
If the Type I and Type II error probabilities at level 0 are equal; i.e., $\alpha_0=\beta_0$, then the unit-threshold likelihood-ratio test reduces to the majority dominance rule. The bounds for the error probabilities in this case and those in the majority dominance rule case are identical. 
\end{Rem}

\subsection{Evenary case}
In the evenary tree case, our results show that with the majority dominance rule, we have
\begin{equation}
\log_2 P_N^{-1}=\Theta(N^{\log_M \lfloor\frac{(M+1)}{2}\rfloor})=\Theta(N^{\log_M \frac{M}{2}}).
\label{equ:upper1}
\end{equation}
This characterizes the explicit convergence rate of the total error probability (c.f. \cite{yash}, in which there is a gap between the upper and lower bounds for $\log_2 P_N^{-1}$). It is evident that the majority dominance rule in this evenary tree case does not achieve the upper bound in~(\ref{equ:upper}). However, the gap between the rates described in (\ref{equ:upper}) and (\ref{equ:upper1}) becomes smaller and more negligible as the degree $M$ of branching grows.

In the case of binary relay trees ($M=2$), the gap is most significant because the total error probability does not change after fusion with the majority dominance rule. In contrast, we have shown in \cite{Zhang} that the likelihood-rate test achieves convergence rate $\sqrt N$. For $M\geq 4$, we have shown that the convergence rate using the Bayesian likelihood-ratio test is at least as good as that using the majority dominance rule.

Now we consider the case where the \emph{alternative majority dominance strategy} (tie is broken alternatively for agents at consecutive levels) is used throughout the tree. In this case we have
\[
\alpha_k=\alpha_{k-1}^M+{M \choose 1}\alpha_{k-1}^{M-1}(1-\alpha_{k-1})+\ldots+{M \choose M/2}\alpha_{k-1}^{M/2}(1-\alpha_{k-1})^{M/2}
\]
and
\[
\alpha_{k+1}=\alpha_{k}^M+{M \choose 1}\alpha_{k}^{M-1}(1-\alpha_{k})+\ldots+{M \choose M/2-1}\alpha_{k}^{M/2+1}(1-\alpha_{k})^{M/2-1}
\]
Using Lemma 1, it is easy to show that 
\begin{align}
1\leq\frac{\alpha_k}{\alpha_{k-1}^{M/2}}\leq {M \choose M/2} \text{  and  }
1\leq\frac{\alpha_{k+1}}{\alpha_{k}^{M/2+1}}\leq {M \choose M/2-1}.
\label{eq:ineq2}
\end{align}

\begin{Theorem}
Consider an $M$-ary relay tree, where $M$ is an even integer, and let $\lambda_M=M/2$. Suppose that we apply the alternative majority dominance strategy. Then, for even $k$ we have
\[
\lambda_M^{k/2} \left(\lambda_M+1\right)^{k/2} \left(\log_2 \alpha_0^{-1}-\log_2 {M\choose \lambda_M}\right) \leq \log_2 \alpha_k^{-1} \leq \lambda_M^{k/2}\left(\lambda_M+1\right)^{k/2}  \log_2 \alpha_0^{-1}.
\]
\label{thmn}
\end{Theorem}
\begin{IEEEproof} The case where $M=2$ is easy to show using the recursion for $\alpha_k$ and the proof is omitted. Now let us consider the case where $M\geq 4$. From the inequalities in \eqref{eq:ineq2}, we have
\[
\alpha_{k+1}=c_k\alpha_k^{\lambda_M+1}=c_k c_{k-1}^{\lambda_M}\alpha_k^{\lambda_M(\lambda_M+1)},
\]
where $c_{k-1} \text{ and } c_k\in \left[1,{M\choose M/2}\right]$. From these we obtain
\[
\alpha_k=c_{k-1}c_{k-2}^{\lambda_M}c_{k-3}^{\lambda_M(\lambda_M+1)} \ldots c_0^{{\lambda_M}^{k/2}(\lambda_M+1)^{k/2-1}} \alpha_0^{{\lambda_M}^{k/2}(\lambda_M+1)^{k/2}},
\]
where $c_i\in \left[1,{M\choose M/2}\right]$ for all $i$.
\begin{align*}
\log_2 \alpha_k^{-1} = &-\log_2 c_{k-1}-\ldots-{\lambda_M}^{k/2}(\lambda_M+1)^{k/2-1} \log_2 c_0 + {\lambda_M}^{k/2}(\lambda_M+1)^{k/2} \log_2 \alpha_0^{-1}.
\end{align*}
Since $\log_2 c_i \in \left[0, \log_2{M\choose M/2}  \right]$, we have
$\log_2 \alpha_k^{-1}\leq {\lambda_M}^{k/2}(\lambda_M+1)^{k/2} \log_2 \alpha_0^{-1}.$
Moreover, we have $\log_2 c_i \leq \log_2{M\choose M/2}$. Hence,
\begin{align}
\nonumber
\label{eq:b}
\log_2 \alpha_k^{-1} \geq &-\log_2{M\choose \lambda_M}(1+\lambda_M+\lambda_M(\lambda_M+1)+\ldots+\lambda_M^{k/2}(\lambda_M+1)^{k/2-1})\\
&+{\lambda_M}^{k/2}(\lambda_M+1)^{k/2} \log_2 \alpha_0^{-1}.
\end{align}
Next we use induction to show that \begin{align}
1+\lambda_M+\lambda_M(\lambda_M+1)+\ldots+\lambda_M^{k/2}(\lambda_M+1)^{k/2-1} \leq \lambda_M^{k/2}(\lambda_M+1)^{k/2}.
\label{eq:dd}
\end{align} 
Suppose that $k=2$. Then, we have $1+\lambda_M\leq \lambda_M(\lambda_M+1)$, which holds because $\lambda_M\geq 2$. Suppose that \eqref{eq:dd} holds when $k=k_0$. We wish to show that it also holds when $k=k_0+1$, in which case we have
\begin{align*}
&1+\lambda_M+\ldots+\lambda_M^{k_0/2}(\lambda_M+1)^{k_0/2-1} +\lambda_M^{k_0/2}(\lambda_M+1)^{k_0/2} +\lambda_M^{k_0/2+1}(\lambda_M+1)^{k_0/2}  \\
&\leq 2\lambda_M^{k_0/2}(\lambda_M+1)^{k_0/2} +\lambda_M^{k_0/2+1}(\lambda_M+1)^{k_0/2} \\
 &\leq 2\lambda_M^{k_0/2+1}(\lambda_M+1)^{k_0/2} \leq \lambda_M^{k_0/2+1}(\lambda_M+1)^{k_0/2+1}.
\end{align*}
Therefore, we have proved \eqref{eq:dd}. Substituting this result in \eqref{eq:b}, we obtain the desired lower bound.
\end{IEEEproof}

The bounds for $\log_2 \beta_k^{-1}$ are similar and they are omitted for brevity. 

\begin{Cor} Let $P_{F,N}$ be the Type I error probability at the root of an $M$-ary relay tree, where $M$ is an even integer. Suppose that we apply the alternative majority dominance strategy. Then, we have
\begin{align*}
N^{\log_M \sqrt{M(M+2)}/2} \left(\log_2 \alpha_0^{-1}-\log_2 {M\choose \lambda_M}\right) \leq
\log_2 P_{F,N}^{-1}
\leq N^{\log_M \sqrt{M(M+2)}/2} \log_2 \alpha_0^{-1}.
\end{align*}
\end{Cor}

\begin{Cor} Let $P_N$ be the total error probability at the root of an $M$-ary relay tree, where $M$ is an even integer. Suppose that we apply the alternative majority dominance strategy. Then, we have 
$\log_2 P_{F,N}^{-1}=\Theta(N^{\log_M \sqrt{M(M+2)}/2})$ and $\log_2 P_{N}^{-1}=\Theta(N^{\log_M \sqrt{M(M+2)}/2}).$

\end{Cor}

Note that when $M=2$, $\log_2 P_{N}^{-1}=\Theta(\sqrt N)$. Therefore, the decay rate with this strategy is identical with that using the Bayesian likelihood-ratio test. This is not surprising because we show in \cite{Zhang} that the Bayesian likelihood-ratio test is essentially either `AND' rule or `OR' rule depending on the values of the Type I and II error probabilities. We also show that the same rule will repeat no more than two consecutive times. Therefore, the decay rate in this case is the same as that using the alternative majority dominance strategy.

For the case where $M\geq 4$, suppose that $\alpha_0$ and $\beta_0$ are sufficiently small and their difference is also sufficiently small. Then, it is easy to show that the Bayesian likelihood-ratio test is majority dominance rule with tie-breaking given by the values of the Type I and II error probabilities. Moreover, we can show that the same tie-breaking will repeat no more than two consecutive times. In this case, the error probability decays as  $\Theta(N^{\log_M \sqrt{M(M+2)}/2}).$

Recall that the upper bound for the decay rate of the total error probability with all combinations of fusion rules is $O(N^{\log_M \frac{M+1}{2}})$, which involves an arithmetic mean of $M+2$ and $M$. In contrast, the decay rate using the alternative majority dominance strategy and Bayesian likelihood-ratio test involves the geometric mean of $M+2$ and $M$, which means that these two strategies are almost asymptotic optimal, especially when $M$ is large. 

\begin{figure}[htbp]
\centering
\includegraphics[width=4.2in]{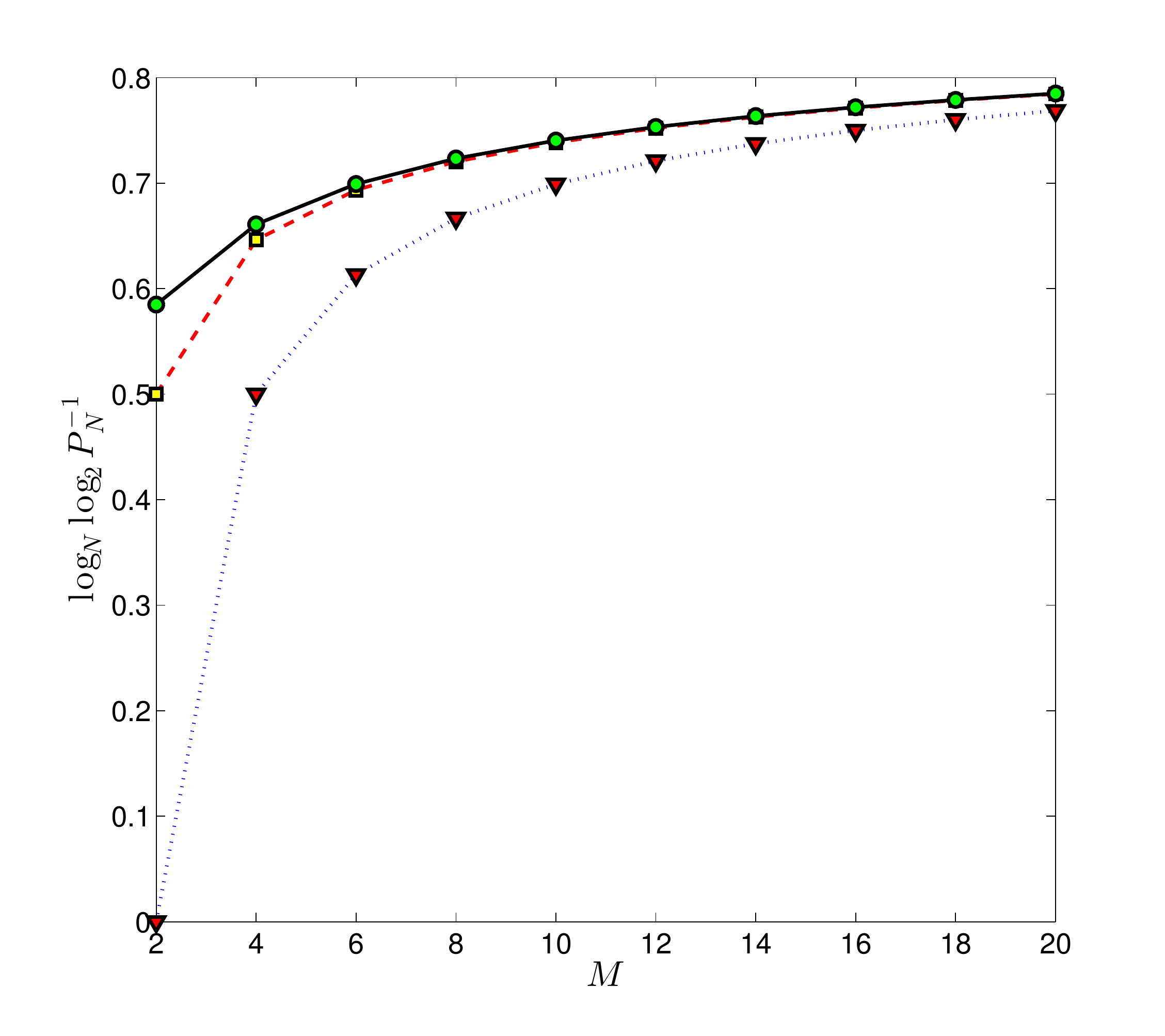}
\vspace{-0.3in}
\caption{Plot of error exponents versus the spanning factor $M$. Dashed (red) line represents the alternative majority dominance strategy. Dotted (blue) line represents the majority dominance rule with random tie-breaking. Solid (black) line represents the exponent in \eqref{equ:upper}. }
\label{fig:comp}
\end{figure}

In addition, the rate using the alternative majority dominance strategy is better comparing to the random tie-breaking case. 
For illustration purposes, in Fig. \ref{fig:comp} we plot the exponent for the decay rate of the total error probability versus the spanning factor $M$ in these two cases. For comparison purposes, we also plot the exponent in the  upper bound \eqref{equ:upper}. We can see from Fig. \ref{fig:comp} that alternative majority dominance strategy achieves a larger exponent than that of the majority dominance rule with random tie-breaking. Moreover, the gap between the exponents in the alternative majority dominance strategy case and the upper bound \eqref{equ:upper} is small and almost negligible.

\section{Non-binary Message Alphabets}
\label{section6}
In the previous sections, each agent in the tree is only allowed to pass a binary message to its supervising agent at the next level. A natural question is, what if each agent can transmit a `richer' message? In this section, we provide a message-passing scheme that allows general message alphabet of size $\mathcal D$ (non-binary). We call this $M$-ary relay tree with message alphabet size $\mathcal D$ an $(M, \mathcal D)$-tree. We have studied the convergence rates of $(M,2)$-trees by investigating how fast the total error probability decays to 0. What about the convergence rate when $\mathcal D$ is an arbitrary finite integer?

We denote by $u^k_o$ the output message for each agent at the $k$th level after fusing $M$ input messages $\textbf{u}_i^{k-1}=\{u^{k-1}_1,u^{k-1}_2,\ldots, u^{k-1}_M\}$ from its child agents at the $(k-1)$th level, where $u^{k-1}_t\in \{0,1,\ldots,\mathcal D\}$ for all $t\in\{1,2,\ldots,M\}$.

\emph{Case I}:
First, we consider an $(M,\mathcal D)$-tree with height $k_0$, in which there are $M^{k_0}$ leaf agents, and the message alphabet size is sufficiently large; more precisely,
\begin{equation}
\mathcal D \geq M^{k_0-1}+1.
\label{equ:d}
\end{equation}

For our analysis, we need the following terminology: 

\emph{Definition}: Given a nonleaf agent in the tree, a \emph{subtree leaf} of this agent is any leaf agent of the subtree rooted at the agent. An \emph{affirmative subtree leaf} is any subtree leaf that sends a message of `1' upward.

Suppose that each leaf agent still generates a binary message $u_o^0\in \{0,1\}$ and sends it upward to its parent agent. Moreover, each intermediate agent simply sums up the messages it receives from its immediate child agents and sends the summation to its parent agent; that is,
$u^{k}_o=\sum^{M}_{t=1}  u^{k-1}_t.$
Then we can show that the output message for each agent at the $k$th level is an integer from $\{0,1,\ldots,M^k\}$ for all $k\in\{0,1,\ldots,k_0-1\}$. Moreover, this message essentially represents the number of its affirmative subtree leaf.

Because of inequality (\ref{equ:d}), at each level $k$ in the tree, the message alphabet size $\mathcal D$ is large enough to represent all possible values of $u_o^k$ ($k\in\{0,\ldots,k_0-1\}$). In particular, the root (at level $k_0$) knows the number of its affirmative subtree leaves. In this case, the convergence rate is the same as that of the star configuration, where each leaf agent sends a binary message to the root directly. Recall that in the star configurations, the total error probability decays exponentially fast to 0.

\begin{figure}[htbp]
\centering
\includegraphics[width=3.2in]{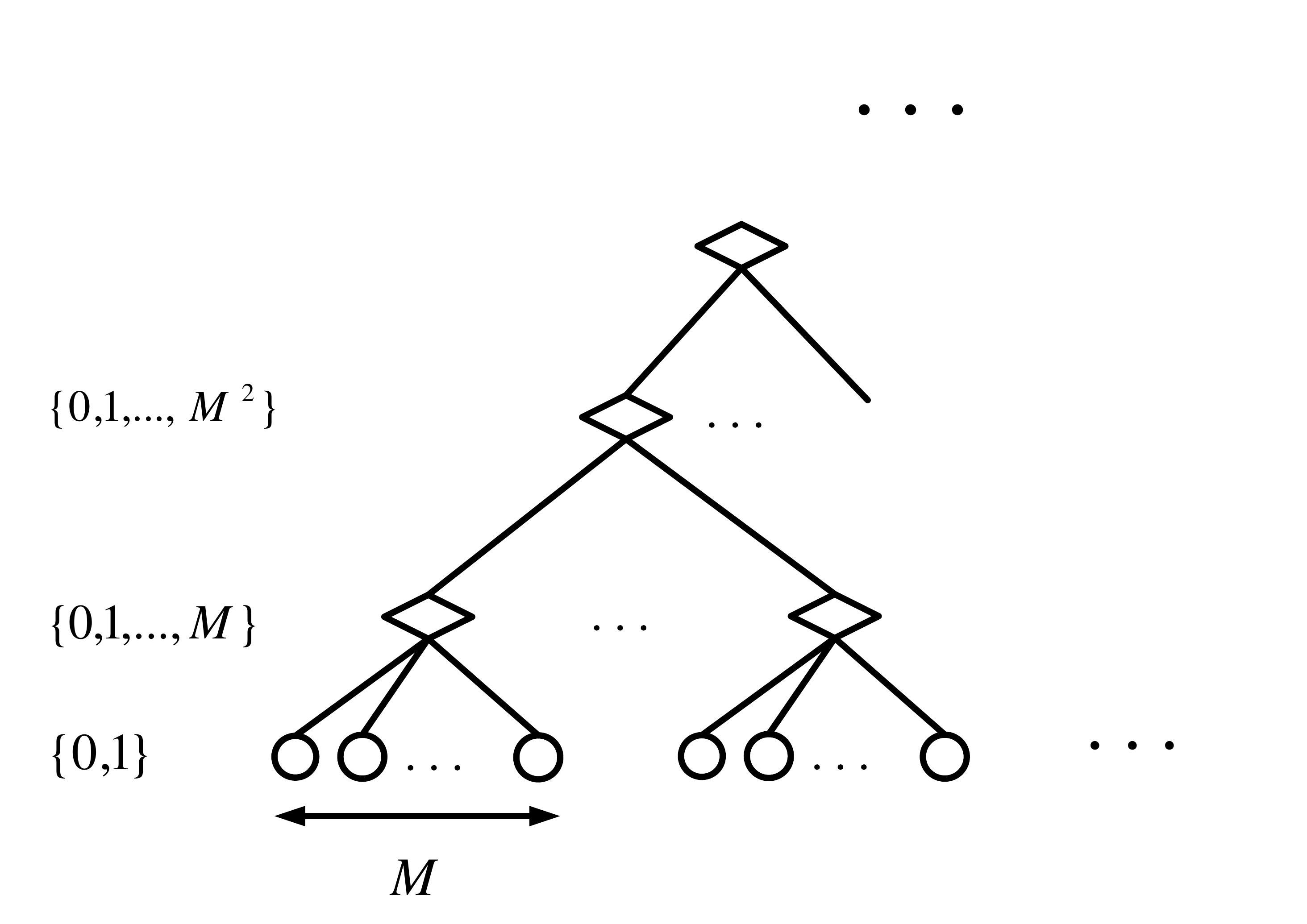}
\vspace{-0.2in}
\caption{A message-passing scheme for non-binary message alphabets in an $M$-ary relay tree.}
\label{fig:a}
\end{figure}

\emph{Case II}: 
We now consider the case where the tree height is very large; i.e., \eqref{equ:d} does not hold. As shown in Fig.~\ref{fig:a}, we apply the scheme described in Case I; that is, the leaf agents send binary compressions of their measurements upward to their parent agents. Moreover, each intermediate agent simply sends the sum of the messages received to its parent agent; i.e.,
\begin{equation}
u^{k}_o=\sum^{M}_{t=1}  u^{k-1}_t.
\label{equ:rule}
\end{equation}
From the assumption of large tree height, it is easy to see that the message alphabet size is not large enough for all the relay agents to use the fusion rule described in (\ref{equ:rule}).
With some abuse of notation, we let $k_0$ to be the integer $k_0=\lfloor\log_M (\mathcal D-1)\rfloor+1$ (here, $k_0$ is not the height of the tree; it is strictly less than the height). Note that
$M^{k_0-1}+1\leq\mathcal D < M^{k_0}+1.$

From the previous analysis, we can see that with this scheme, each agent at the $k_0$th level knows the number of its affirmative subtree leaves. Therefore, it is equivalent to consider the case where each agent at level $k_0$ connects to its $M^{k_0}$ subtree leaves directly (all the intermediate agents in the subtree can be ignored).
However, we cannot use the fusion rule described in (\ref{equ:rule}) for the agents at $k_0$th level to generate the output messages because the message alphabet size is not large enough. Hence, we let each agent at level $k_0$ aggregate the $M^{k_0}$ binary messages from its subtree leaves into a new binary message (using some fusion rule). By doing so, the output message from each agent at the $k_0$th level is binary again. Henceforth, we can simply apply the fusion rule (\ref{equ:rule}) and repeat this process throughout the tree, culminating at the root. We now provide an upper bound for the asymptotic decay rate in this case.

\begin{Theorem} The convergence rate of the total error probability for an $(M, \mathcal D)$-tree is equal to that for an $(M^{k_0}, 2)$-tree, where $k_0=\lfloor\log_M (\mathcal D-1)\rfloor+1$. In particular, let $P_N$ be the total error probability at the root for an $(M, \mathcal D)$-tree. With any combination of fusion rules at level $\ell k_0$, $\ell=1,2,\ldots,$ we have
$\log_2 P_N^{-1}=O\left(N^\rho\right),$
where
\begin{equation*}
\rho:=
\frac{\ln(M^{k_0}+1)}{\ln M^{k_0}}-\frac{\log_M 2}{k_0}.
\end{equation*}

\label{thm5'}
\end{Theorem}

\begin{IEEEproof}
Consider an $(M, \mathcal D)$-tree with the scheme described above. It is easy to see that equivalently we can consider a tree where the leaf agents connect to the agents at the $k_0$th level directly. In addition, because of the recursive strategy applied throughout the tree, it suffices to consider the tree where the agents at the $\ell k_0$th level connect to the agents at the $(\ell+1)k_0$th level directly for all non-negative integers $\ell$. Therefore, the convergence rate of an $(M, \mathcal D)$-tree is equal to that of the corresponding $(M^{k_0},2)$-tree.

In the asymptotic regime, the decay rate in $(M,2)$-trees is bounded above as follows \cite{yash}:
\begin{equation*}
\log_2 P_N^{-1}=O(N^{\log_M \frac{(M+1)}{2}}).
\end{equation*}
Therefore, the decay rate for $(M^{k_0}, 2)$-trees is also bounded above as
\[
\log_2 P_N^{-1}=O(N^{\log_{M^{k_0}} \frac{(M^{k_0}+1)}{2}}),
\]
which upon simplification gives the desired result.
\end{IEEEproof}

Suppose that each agent at level $\ell k_0$ for all $\ell$ uses the majority dominance rule. Then, we can derive the convergence rate for the total error probability as follows.

\begin{Theorem}
Consider $(M, \mathcal D)$-trees where the majority dominance rule is used. Let $k_0=\lfloor\log_M (\mathcal D-1)\rfloor+1$. 
We have
$\log_2 P_N^{-1}=\Theta\left(N^\varrho\right),$
where
\begin{equation*}
\varrho:=\left\{\begin{array}{l l}
\frac{\ln(M^{k_0}+1)}{\ln M^{k_0}}-\frac{\log_M 2}{k_0}, & \quad  \text{ if $M$ is odd},\\
1-\frac{\log_M 2}{k_0}, & \quad  \text{ if $M$ is even}. \end{array}\right.
\end{equation*}
\label{thm5}
\end{Theorem}
\begin{IEEEproof}
By Theorem \ref{thm5'}, the performance of $(M, \mathcal D)$-trees is equal to that of $(M^{k_0}, 2)$-trees, where $k_0=\lfloor\log_M (\mathcal D-1)\rfloor+1$. For the asymptotic rate, we have
\[
\log_2 P_N^{-1}=\Theta(N^{\log_{M^{k_0}} \left\lfloor \frac{M^{k_0}+1}{2}\right\rfloor}),
\]
which upon simplification gives the desired result.
\end{IEEEproof}

\begin{Rem}
Notice that $\lim_{M\to \infty}\ln(M^{k_0}+1)/\ln M^{k_0}=1$, which means that the even and odd cases in the expression for $\varrho$ are similar when $M$ is large. 
\end{Rem}
\begin{Rem}
From Theorem \ref{thm5}, we can see that with larger message alphabet size, the total error probability decays more quickly. However, the change in the decay exponent is not significant because $k_0$ depends on $\mathcal D$ logarithmically. Furthermore, if $M$ is large, then the change in the performance is less sensitive to the increase in $\mathcal D$. 
\end{Rem}

\begin{Rem}
Comparing the results in Theorems \ref{thm5'} and \ref{thm5}, we can see that the majority dominance rule achieves the optimal exponent in the oddary case and it almost achieves the optimal exponent in the evenary case.
\end{Rem}

For the Bayesian likelihood-ratio test, we have the following result.
\begin{Theorem} The convergence rate using the likelihood-ratio test is at least as good as that using the majority dominance rule; i.e.,
$\log_2 \mathds P_N^{-1}=\Omega\left(N^\varrho\right).$
\end{Theorem}

In the case where $M$ is even, we can derive the decay rate using the alternative majority dominance strategy. 
\begin{Theorem} The convergence rate using the alternative majority dominance strategy is 
$\log_2 P_N^{-1}=\Omega\left(N^\sigma \right),$ where 
\[
\sigma=\frac{1}{2}\left( 1+\frac{\ln (M^{k_0}+2)}{\ln M^{k_0}} \right) - \frac{\log_M 2}{k_0}.
\]
\end{Theorem}

Theorem 8 and 9 follow by applying the same arguments as those made in proofs of Corollary 4 and Theorem 6 and the proofs are omitted for brevity.

The message-passing scheme provided here requires message alphabets with maximum size $\mathcal D$. However, most of the agents use much `smaller' messages. For example, the leaf agents generate binary messages. It is interesting to characterize the \emph{average} message size used in our scheme.
Because of the recursive strategy, it suffices to calculate the average message size in a subtree with height $k_0-1$ since the message sizes in our scheme repeat every $k_0$ levels.
The message size (in bits) for agents at level $t\in \{0,1,\ldots,k_0-1\}$ is $\log_2(M^t+1)$ and the number of agents at level $t$ is $M^{k_0-t}$. Therefore,
the average size $\overline{b}(k_0)$ in bits used in our scheme is
\begin{align*}
\overline{b}(k_0)=&\frac{M^{k_0}+\ldots+M \log_2 (M^{k_0-1}+1)}{M^{k_0}+M^{k_0-1}+\ldots+M}
=\frac{\sum_{t=0}^{k_0-1} M^{k_0-t}\log_2 (M^{t}+1)}{\sum_{t=0}^{k_0-1} M^{t+1}}
\end{align*}
We have
\[
\log_2 (M^t+1)>\log_2 M^t=t\log_2 M
\]
and
\[
\log_2 (M^t+1)< \log_2(2M^t)=1+t\log_2 M
\]
for all $t\geq 1$.
Therefore, the average size in bits is lower bounded as
\begin{align*}
\overline{b}(k_0)> & \frac{M^{k_0}+M^{k_0-1}\log_2 M+\ldots+M (k_0-1)\log_2 M}{M^{k_0}+M^{k_0-1}+\ldots+M}\\
=&\frac{M^{k_0}}{M^{k_0}+M^{k_0-1}+\ldots+M}  +\frac{\log_2 M (M^2(M^{k_0-1}-1)-M(M-1)(k_0-1))}{(M^{k_0}+M^{k_0-1}+\ldots+M)(M-1)^2}\\
=& \frac{M^{k_0}-M^{k_0-1}}{M^{k_0}-1}+\frac{M\log_2 M}{M-1}\frac{M^{k_0-1}-1-M(M-1)(k_0-1)}{M^{k_0}-1}.
\end{align*}
In addition, it is upper bounded as
\begin{align*}
\overline{b}(k_0)&<1+\frac{M\log_2 M}{M-1}\frac{M^{k_0-1}-1-M(M-1)(k_0-1)}{M^{k_0}-1} \leq 1+\frac{\log_2 M}{M-1}.
\end{align*}

\begin{figure}[htbp]
\begin{center}
\begin{tabular}{cc}
\includegraphics[width=2.6in]{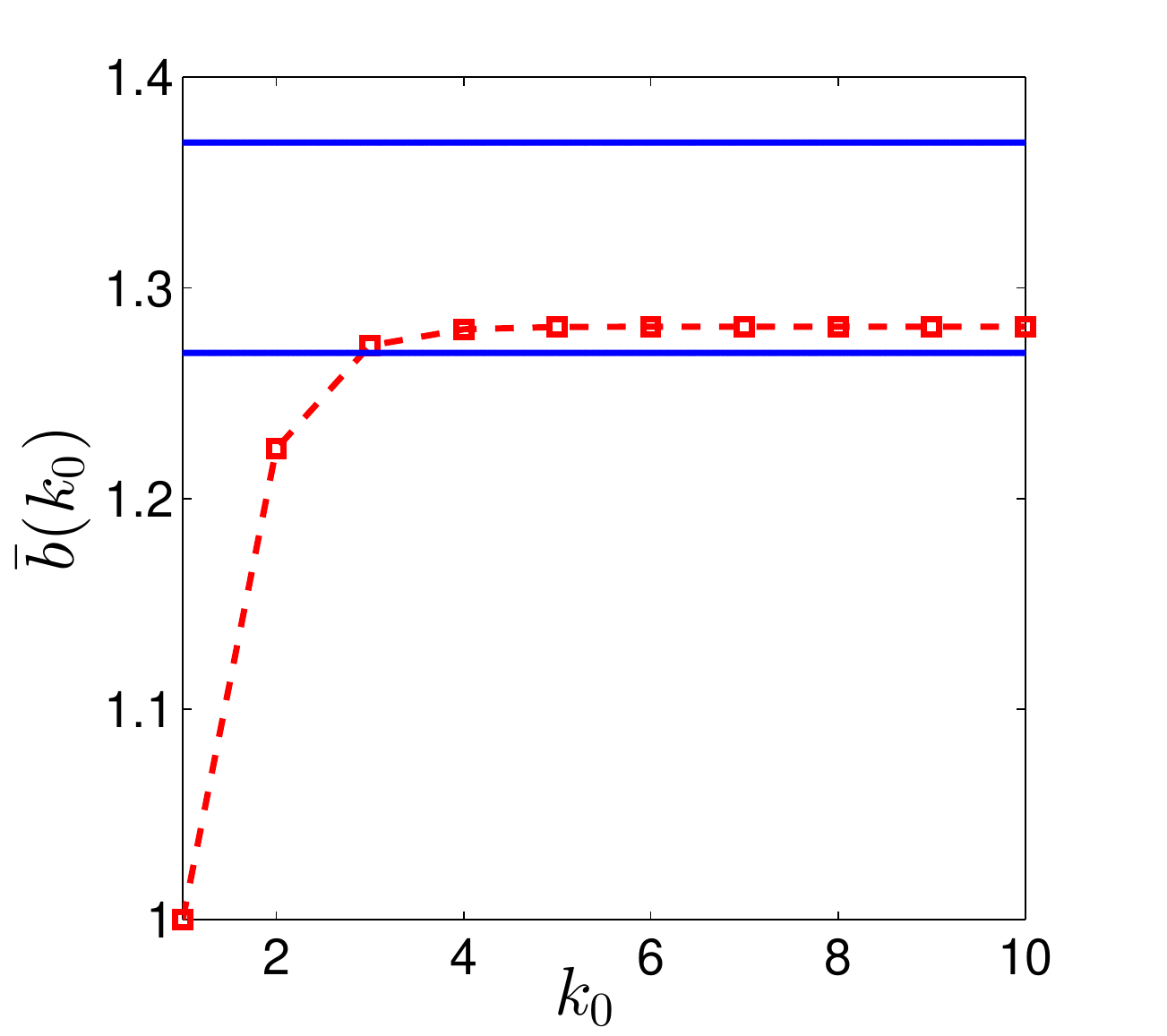} & \includegraphics[width=2.6in]{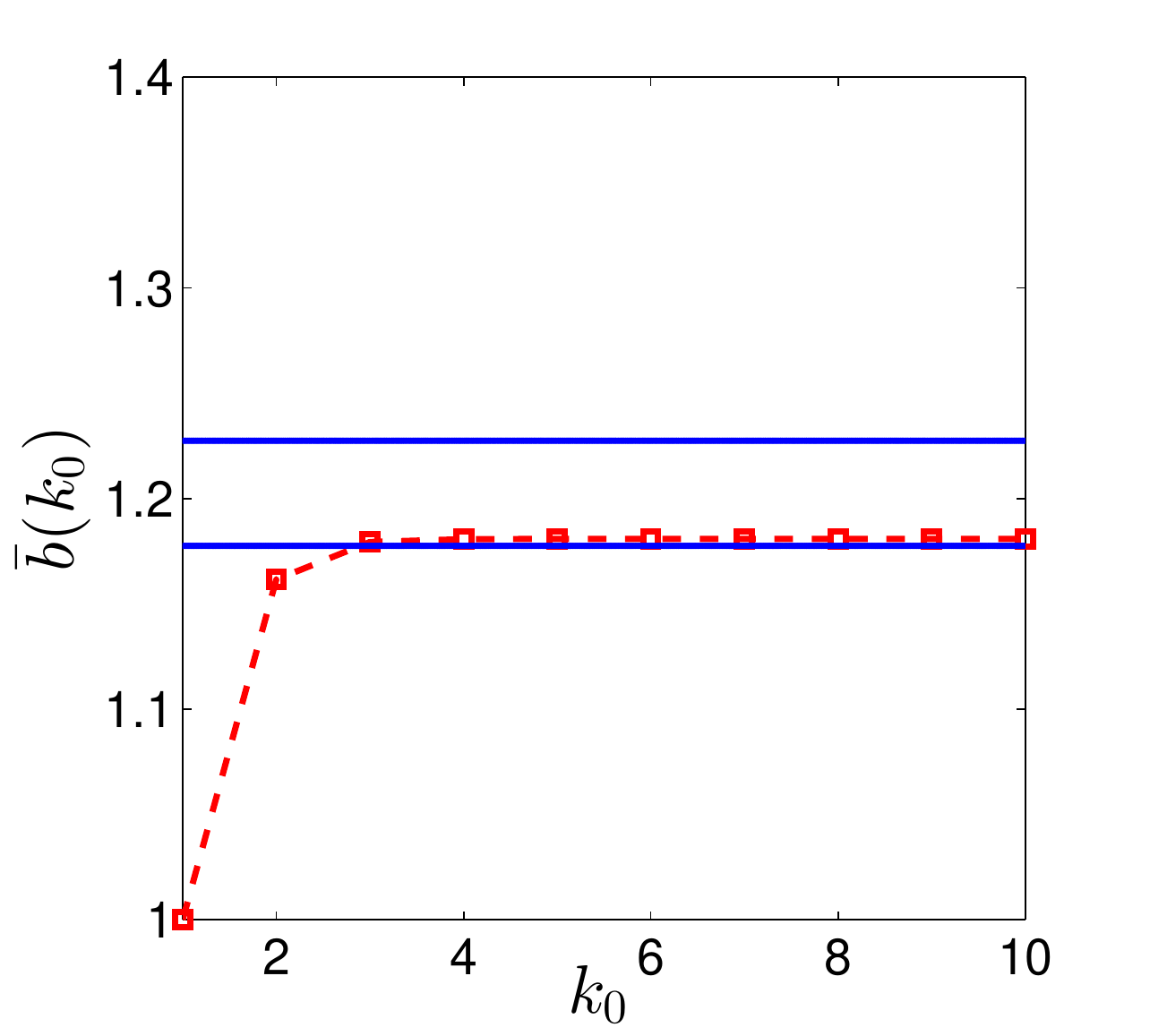} \\
(a)& (b)\\
\end{tabular}
\end{center}
\caption{(a) Average message size (dashed red line) in $M=10$ case. (b) Average message size (dashed red line) in $M=20$ case. The blue lines represent the bounds in (\ref{equ:bound}).}
\label{fig:size}
\end{figure}

Recall that, with sufficiently large $k_0$, the error probability convergence rates are close to exponential. However, from the above bounds the average message size in terms of bits in our scheme is still very small, specifically for sufficiently large $k_0$ we have
\begin{align}
1+\frac{\log_2 M}{M-1}-\frac{1}{M}\leq\overline{b}(k_0)\leq 1+\frac{\log_2 M}{M-1}.
\label{equ:bound}
\end{align}
Fig. \ref{fig:size} shows plots of the average message sizes $\overline{b}(k_0)$ versus $k_0$ in the $M=10$ and 20 cases. Note that as $M$ increases, the average message size becomes smaller and the bounds in (\ref{equ:bound}) become tighter.

\section{Concluding Remarks}
\label{section7}

We have studied the social learning problem in the context of $M$-ary relay trees. We have analyzed the step-wise reductions of the Type I and Type II error probabilities and derived upper and lower bounds for each error probability at the root as explicit functions of $N$, which characterize the convergence rates for Type I, Type II, and the total error probabilities. We have shown that the majority dominance rule is not better than the Bayesian likelihood-ratio test in terms of convergence rate. We have studied the convergence rate using the alternative majority dominance strategy, which in turn shows that the majority dominance rule with random tie-breaking is suboptimal in the case where $M$ is even. Last, we have provided a message-passing scheme which increases the convergence rate of the total error probability. We have shown quantitatively how the convergence rate varies with respect to the message alphabet sizes. This scheme is very efficient in terms of the average message size used for communication.

Many interesting questions remain. Social networks usually involve very complex topologies. For example, the degree of branching may vary among different agents in the network. The convergence rate analysis for general complex structures is still wide open.
Another question involves the assumption that the agent measurements are conditionally independent. It is of interest to study the scenario where these agent measurements are correlated. This scenario has been studied in the star configuration \cite{Ve}--\nocite{Dai}\cite{Hao1} but not in any other structures yet.
Yet another question is related with the assumption that the communications and agents are perfectly reliable. We would like to study the rate of learning in cases where communications and agents are non-ideal \cite{Zhang2}.
\vspace{-0.1in}

\appendices
\section{Proof of Proposition \ref{prop2}}
\begin{IEEEproof}
We consider the ratio of $\alpha_{k+1}$ and $\alpha_k^{M/2}$:
\[
\frac{\alpha_{k+1}}{\alpha_k^{M/2}}=\alpha_k^{M/2}+{M \choose 1} \alpha_k^{(M-2)/2}(1-\alpha_k)+\ldots+\frac{1}{2}{M\choose M/2} (1-\alpha_k)^{M/2}.
\]
First, we show the lower bound of the ratio. We know that
\begin{align*}
(\alpha_k+1-\alpha_k)^{M/2}&=\alpha_k^{M/2}+{M/2 \choose 1}\alpha_k^{(M-2)/2}(1-\alpha_k)+\ldots+{M/2\choose M/2} (1-\alpha_k)^{M/2}=1
\end{align*}
and
${M \choose k}\geq {M/2 \choose k}$
for all $k=1,2,\ldots, M/2$. Moreover, we have
${M\choose M/2}/2\geq {M/2\choose M/2} =1.$
In consequence, we have
${\alpha_{k+1}}/{\alpha_k^{M/2}}\geq 1.$
Notice that $\alpha_{k+1}/\alpha_k^{M/2} =h_{M/2}^M(\alpha_k)/2 + h_{M/2-1}^M(\alpha_k)/2$. By Lemma 1, the ratio is monotone increasing as $\alpha_k\to 0$.
Hence, we have ${\alpha_{k+1}}/{\alpha_k^{M/2}}\leq \frac{1}{2}{M\choose M/2}.$
\end{IEEEproof}

\section{Proof of Theorem \ref{thm2}}
\begin{IEEEproof}
From the inequalities in Proposition \ref{prop2} been derived, we have
$\alpha_{k+1}=c_k\alpha_k^{M/2}=c_k\alpha_k^{\lambda_M},$ 
where $c_k\in \left[1,{M\choose M/2}/2\right]$. From these we obtain
\[
\alpha_k=c_{k-1}c_{k-2}^{\lambda_M} \ldots c_0^{{\lambda_M}^{k-1}} \alpha_0^{{\lambda_M}^k},
\]
where $c_i\in \left[1,{M\choose M/2}/2\right]$ for all $i$.
\begin{align*}
\log_2 \alpha_k^{-1} = &-\log_2 c_{k-1}- \lambda_M \log_2 c_{k-2} -\ldots-\lambda_M^{k-1} \log_2 c_0 + \lambda_M^{k} \log_2 \alpha_0^{-1}.
\end{align*}
Since $\log_2 c_i \in \left[0, \log_2{M\choose M/2} -1 \right]$, we have
$\log_2 \alpha_k^{-1}\leq \lambda_M^{k} \log_2 \alpha_0^{-1}.$
Moreover, we obtain
\begin{align*}
\log_2 \alpha_k^{-1} \geq &-\log_2{M\choose M/2} -\lambda_M \log_2{M\choose M/2} -\ldots-\lambda_M^{k-1} \log_2{M\choose M/2} + \lambda_M^{k} \log_2 \alpha_0^{-1} \\
 \geq& \lambda_M^k \left(\log_2 \alpha_0^{-1}-\log_2{M\choose M/2}  \right).
\end{align*}
\end{IEEEproof}

%
%
%

\ifCLASSOPTIONcaptionsoff
  \newpage
\fi

\bibliographystyle{IEEEbib}

\end{document}